# Fabrication of hydrogen-bonded metal inorganic-organic complex glasses by ligand-tuning approach

Tianzhao Xu[1], Zhencai Li[1*], Jia-Xin Wu[2], Zihao Wang[3], Hanmeng Zhang[4], Huotian Zhang[5], Lars R. Jensen[6], Kenji Shinozaki[7], Feng Gao[5], Haomiao Zhu[3], Ivan Hung[8], Zhehong Gan[8], Jinjun Ren[4], Zheng Yin[2], Ming-Hua Zeng[2*], Yuanzheng Yue[1*]

[1]Department of Chemistry and Bioscience, Aalborg University, DK-9220 Aalborg, Denmark

[2]School of Chemistry and Pharmaceutical Sciences, State Key Laboratory for Chemistry and Molecular Engineering of Medicinal Resources, Guangxi Normal University, Guilin 541004, P. R. China

[3]Xiamen Research Center of Rare Earth Materials, Haixi Institutes, Chinese Academy of Sciences, Xiamen 361021, China

[4]Shanghai Institute of Optics and Fine Mechanics, Chinese Academy of Sciences, Shanghai 201800, China

[5]Department of Physics, Chemistry, and Biology (IFM), Linköping University, Linköping 583 30, Sweden

[6]Department of Materials and Production, Aalborg University, DK-9220 Aalborg, Denmark

[7]Research Institute of Core Technology for Materials Innovation, National Institute of Advanced Industrial Science and Technology (AIST), Ikeda, Osaka 563-8577, Japan

[8]National High Magnetic Field Laboratory, FL 32310, USA

*Corresponding author. E-mail: zhli@bio.aau.dk, zmh@mailbox.gxnu.edu.cn, yy@bio.aau.dk

# Abstract

Metal inorganic-organic complex (MIOC) crystals are a new category of hybrid glass formers. However, the glass-forming compositions of MIOC crystals are limited due to lack of both a general design principle for such compositions and a deep understanding of the structure and formation mechanism for MIOC glasses. This work reports a general approach for synthesizing glass-forming MIOC crystals. In detail, the principle of this approach is based on the creation of hydrogen-bonded structural network by substituting acid anions for imidazole or benzimidazole ligands in the tetrahedral units of zeolitic imidazolate framework crystals. By tuning the metal centers, anions, and organic ligands of MIOCs, supramolecular unit structures can be designed to construct supramolecular networks and thereby enable property modulation. Furthermore, mixed-ligand synthesis yielded a mixed-crystal system in which the glass-transition temperature ($T_g$) can be linearly tuned from 282 K to 360 K through gradual substitution of benzimidazole for imidazole. Interestingly, upon vitrification, MIOCs were observed to undergo reorganization of hydrogen-bonded networks, with retention of tetrahedral units, short-range disorder, and the freezing of multiple conformations. This work offers a new strategy to systematically expand the glass-forming compositional range of MIOCs and to develop functional MIOC glasses.

# Introduction

Metal complex (MC) crystals, composed of metal ions or atoms coordinated with organic or inorganic ligands, represent the cornerstone of coordination chemistry and have been studied for over a century[1]. Recently, melt-quenched MC crystals have attracted increasing attention, leading to the emergence of distinct branches such as metal-organic framework (MOF) glasses, coordination polymer (CP) glasses, and metal inorganic-organic complex (MIOC) glasses, which exhibit three- to zero-dimensional network structures[2-4]. Due to their structural diversity, high tunability, and designability, MC crystals can be potentially applied in many fields, including gas separation[5, 6], lithium-ion batteries[7], and luminescent host materials[4]. Compared to MC crystals, the upscaling applications of melt-quenched MC glasses are enabled by their bulk nature and enhanced processability[3, 8]. However, the preparation of MC glasses is seriously constrained by the ligand decomposition[9] or molecular sublimation[10] upon heating, resulting in a limited number of reported examples to date.

Within MCs, some zeolitic imidazolate frameworks (ZIFs), a subclass of MOFs, exhibit an ultra-high glass-forming ability and a tetrahedral network structure analogous to that of $SiO_2$[11]. According to the literature, various ligands such as solvents[12], phosphates[13], and halides[14] can be introduced into ZIFs to generate CPs or MIOCs. This introduction can partially replace the coordination bonds in ZIFs with weaker interactions (e.g., hydrogen bonds or weak coordination), thereby reducing the melting barrier and enabling the fabrication of bulk MC glasses. Their processability and structural flexibility confer unique properties, offering promising applications in areas such as luminescent hosts[15], room-temperature phosphorescence[16], and gas separation membranes[17]. However, up to now, investigations on MIOC glasses have mostly focused on particular compounds and properties. Their underlying glass-forming mechanisms have not been elucidated through multilevel structural modulation or by tracking structure evolution during vitrification. The lack of detailed structural characterization of MIOC glasses further constrains the exploration of new glass-forming systems. Therefore, it is crucial to advance our understanding of MIOC glasses and develop a universal approach for broadening the composition range of MIOC glasses.

In this work, MIOC crystals were synthesized via a slow evaporation approach by introducing chloride into two types of ZIFs (i.e., ZIF-4 and ZIF-7)[18], specifically $ZnCl_2HIm_2$ (MIOC-4) and $ZnCl_2HbIm_2$ (MIOC-7) (HIm: imidazole; HbIm: benzimidazole). The supramolecular structures of both MIOCs are predominantly assembled via hydrogen bonding, whereas the benzimidazole ligands in MIOC-7 further contribute extensive π-π interactions. Interestingly, a binary MIOC mixed-crystal system was obtained through mixed-ligand synthesis, exhibiting reduced melting points and tunable glass-transition temperatures. The structural evolution of MIOC crystals during melting was characterized by in situ X-ray diffraction (XRD). The structures of both crystalline and glassy MIOCs were investigated by ex situ XRD, Fourier transform infrared (FT-IR) transmittance, Raman spectroscopy, high-energy synchrotron XRD, and solid-state nuclear magnetic resonance (NMR) spectroscopy. The melting point ($T_m$) of the MIOCs and the glass transition temperature ($T_g$) of the MIOC glasses were measured using differential scanning calorimetry (DSC). In addition, the design strategy can be expanded through multilevel modulations of the metal nodes (Cu/Co substituting Zn) and anions ($SCN^-$ substituting Cl⁻). The photonic properties of MIOC-7 crystals and glasses were characterized, revealing the correlation between structural design and material performance. Our work offers deep insights into the vitrification mechanism of MIOC crystals and a new avenue for developing novel functional MIOC glasses.

# Results and discussion

## Characteristics of MIOC crystals and glasses

Two Zn-based MIOCs, MIOC-4 ($ZnCl_2HIm_2$) and MIOC-7 ($ZnCl_2HbIm_2$), are synthesized using the slow evaporation method (see Supporting Information). In the supramolecular structure of MIOCs, the imidazole ligands are primarily involved in hydrogen bonding, whereas the benzimidazole ligands are capable of introducing more π-π interactions, giving rise to potential fluorescence properties. To characterize the crystalline structure, the XRD patterns of the two MIOC crystals are collected and refined, as shown in Figs. 1a, S1 and S2. From the XRD results, MIOC-4 is identified to be a monoclinic phase[19], whereas MIOC-7 is a triclinic phase[20]. To accurately reveal the supramolecular interaction within the crystal structure, DFT (density functional theory) calculation is carried out. All input files for the CP2K program were generated using the Multiwfn program[21-23]. The crystal structures of both MIOCs are assembled and stabilized by intermolecular interactions, including hydrogen bonding and π-π stacking (Figs. 1b and c). Specifically, in MIOC-4, two types of $[ZnCl_2(HIm)_2]$ tetrahedra are interdigitated and linked by N-H···Cl hydrogen bonds to form zigzag chains along the $[10\bar{1}]$ direction. These chains further extend through N-H···Cl and C-H···Cl hydrogen bonds into two-dimensional hydrogen-bonded layers in the ac plane. The weak hydrogen bonds and π-π interactions govern the stacking of the hydrogen-bonded layers into an ABAB arrangement, thereby resulting in the supramolecular structure (Figs. S3 and S4). In MIOC-7, $[ZnCl_2(HbIm)_2]$ tetrahedra are linked by N-H···Cl hydrogen bonds to form one-dimensional chains along the $[100]$ direction. These chains extend through N-H···Cl hydrogen bonds along the b axis into two-dimensional layers, which interdigitate via interlayer C-H···Cl hydrogen bonds and π-π interactions. Finally, the interdigitated layers are stacked through π-π interactions, resulting in the formation of the supramolecular structure (Tables S1 and Figs. S5-7).

Figs. 1d and e show the thermal responses of MIOC-4 and MIOC-7 during DSC upscan, respectively. The endothermic signal observed in the first upscan was ascribed to melting, whereas that in the second upscan was attributed to the glass transition. During this process, the MIOC-7 crystal exhibits a slight mass loss of about 0.5%, attributed to the release of the residual solvent (acetone) trapped within the as-synthesized crystal. No further mass loss is detected over three consecutive upscan-downscan cycles for MIOC-7 (Fig. S8), indicating the total desolvation of acetone during the first upscan. The offset values of the melting peak, i.e., the melting point[24], for MIOC-4 and MIOC-7 are determined to be 440 and 508 K, respectively. The higher $T_m$ of MIOC-7 hints at the following scenario. Due to its larger size, the HbIm ligand may cause a higher degree of steric hindrance, while the interdigitated arrangement provides greater structural stability than the stacked counterpart. These factors suppress the mobility of structural units and hence require a higher temperature to break down the network for melting. The large-sized transparent MIOC-4 and MIOC-7 glasses are obtained by melt-quenching the as-synthesized MIOC crystals at temperatures above their respective $T_m$ in air, as shown in Fig. 1f. No sharp XRD Bragg's peak is observed for the melt-quenched MIOC glasses (Fig. 1a), indicating their amorphous nature. $T_g$ values of MIOC-4 and MIOC-7 glasses are determined to be 282 and 359 K, respectively (Figs. 1d and e). The higher $T_g$ of MIOC-7 is ascribed to the pronounced steric hindrance of the HbIm ligand, which increases the rigidity of the glass network and restricts molecular dynamics.

To further modulate the thermal properties of MIOC glasses, a series of samples with mixed HIm and HbIm ligands was synthesized by varying the ligand ratio. To confirm the composition of the product, solution-state $^1$H NMR spectra of the mixed-ligand samples are collected (Figs. S9-14), showing that the experimental ligand composition in the binary system closely matches the theoretical one. XRD patterns and refinement analysis (Figs. 1g and S15) revealed that all the mixed-ligand compounds are physical mixtures of MIOC-4 and MIOC-7. However, the obtained amount ratios between MIOC-4 and MIOC-7 crystals deviate from the ratios between HIm and HbIm ligands. The observed discrepancy between ligand and crystal structural ratios might be attributable to low-crystallinity phases, mixed-ligand molecules, and the small-sized MIOC crystals in the binary system, which cannot be detected by XRD.

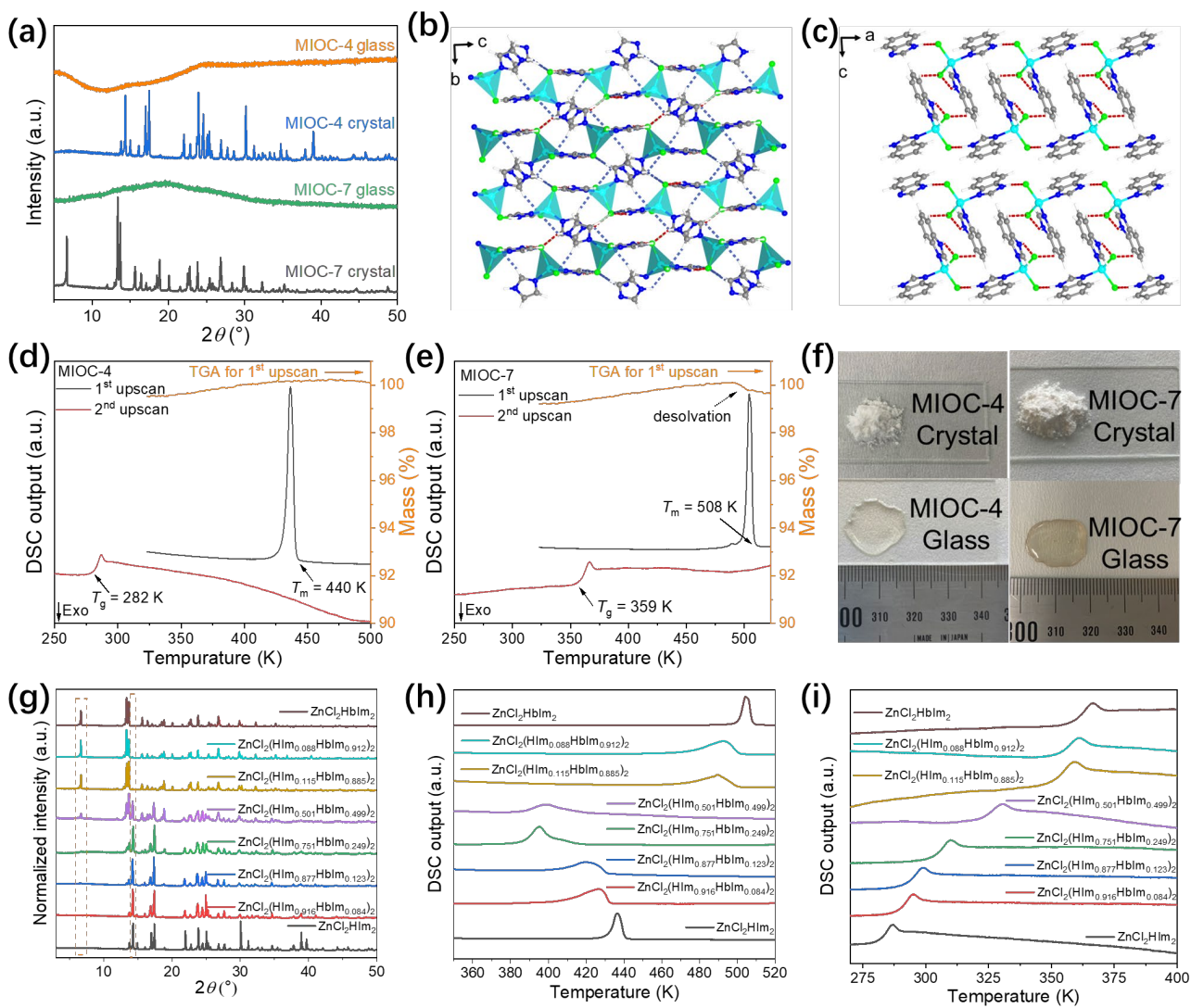

**Figure 1. (a)** XRD patterns of the synthesized MIOC crystals and glasses. **(b)** The crystal structure of MIOC-4. (Noting: blue spheres represent Zn, gray spheres represent C, green spheres represent Cl, white spheres represent H, and dark blue spheres represent N.) **(c)** The crystal structure of MIOC-7. **(d)** DSC and Thermogravimetric analysis (TGA) curves of MIOC-4. **(e)** DSC and TGA curves of MIOC-7. **(f)** The real pictures of the synthesized MIOC-4 and MIOC-7 crystals and the melt-quenched glasses. **(g)** XRD patterns of the binary mixed-crystal system. **(h)** The first DSC upscan of the binary mixed-crystal system. **(i)** The second DSC upscan of the binary mixed-crystal system.

To evaluate the compositional impact on the thermal properties of the binary MIOC system, their $T_g$ and $T_m$ values (Figs. 1h and i, and Table S2) are determined from the DSC upscan curves shown in Figs. S16-21. The $T_g/T_m$ ratio can be employed as an indirect measure of glass-forming ability (GFA), with a higher $T_g/T_m$ ratio corresponding to a high GFA[11]. Figs. S22-24 reveal the correlations of $T_m$, $T_g$, and the $T_g/T_m$ ratio with the molar ratio of $R$ = HbIm/(HIm + HbIm) in the binary MIOC system. It is clearly seen that the $T_g/T_m$ ratio of the binary MIOC system increases by gradually adding HbIm to MIOC-4. In contrast, the $T_g/T_m$ ratio of the binary MIOC system only slightly decreases from 0.71 to 0.70 by introducing a small amount of HIm into MIOC-7. The $T_g/T_m$ ratio of the binary MIOC system is influenced by both the decrease in $T_m$ of the mixed-crystal products and the gradual increase in $T_g$ of the glasses with increasing HbIm content. The reduction in $T_m$ arises from the formation of a mixed-crystal system by the mixed-ligand products[25], while defects such as dislocations, microcrystals, and mixed-ligand units generated by molecular-level mixing further broaden the melting peaks. The linear increase of $T_g$ values with increasing $R$ implies that the replacement of HbIm for HIm results in larger Zn-ligand tetrahedra, thereby enhancing the steric hindrance of the structural network in the glass[11].

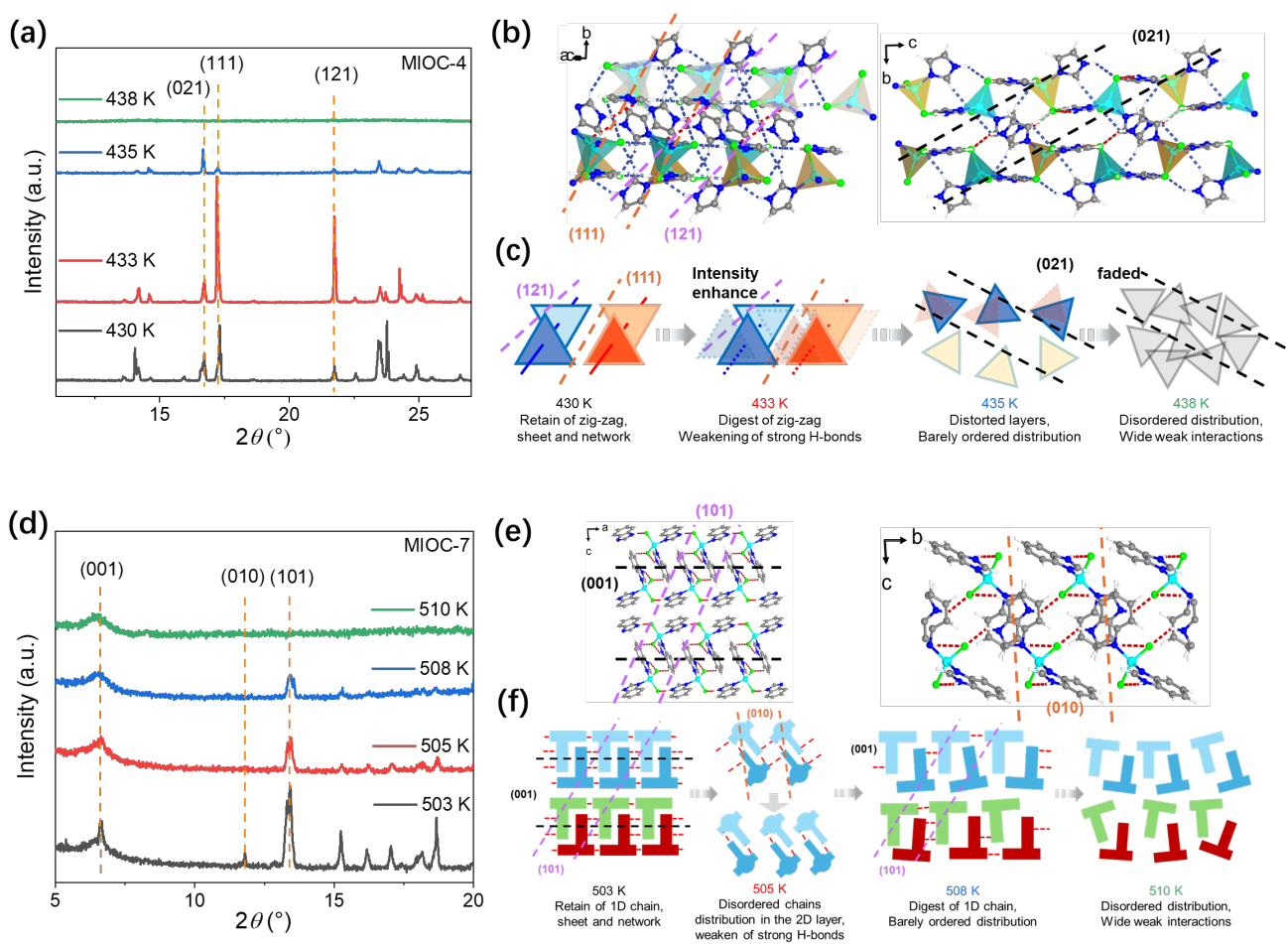

**Figure 2. (a)** Variable temperature XRD patterns of MIOC-4 during the melting process. **(b)** Crystal planes included in the MIOC-4 structural evolution analysis. **(c)** Schematic diagram of the structural evolution of MIOC-4 during melting. **(d)** Variable temperature XRD patterns of MIOC-7 during the melting process. **(e)**

Crystal planes included in the MIOC-7 structural evolution analysis. **(f)** Schematic diagram of the structural evolution of MIOC-7 during melting.

To investigate the origin of both the structural evolution and vitrification of MIOC crystals, an in-situ experiment of XRD for the two MIOC crystals was performed during their melting. Variable-temperature XRD patterns of MIOC-4 and MIOC-7 (Figs. S25 and S26) show that their crystal structures were found to remain intact prior to melting, with no phase transition detected. In MIOC-4, peaks corresponding to the (121) and (111) planes increase in intensity at the early stage of melting and then vanish abruptly, whereas the peak corresponding to the (0 2 1) plane remains well preserved until complete melting (Fig. 2a). In the crystal structure of MIOC-4 (Fig. 2b), the zigzag chains formed by tetrahedra linked through the strongest N-H⋯Cl hydrogen bonds extend within the (121) and (111) planes. Therefore, the observed evolutions of the peak intensity can be attributed to changes in the chains and their stacking arrangements. In contrast, the (021) plane intersects the zigzag chains and is associated with the weakest hydrogen bonds and π-π interactions between adjacent chains. At the initial stage of melting (433 K), the weakening of strong hydrogen bonds induces distortion and slippage of the zigzag chains, causing heavy atoms (such as Zn) to occupy the planes and thereby enhancing the intensity of the chain-related reflections. Subsequently (435 K), the N-H⋯Cl bonds break and the zigzag chains rapidly disintegrate, leaving only weak interactions between adjacent molecules to maintain the crystal structure. Further heating (438 K) leads to complete disorder of intermolecular correlations, resulting in the formation of a melt (Fig. 2c).

In MIOC-7, the intensities of all diffraction peaks decrease with increasing temperature, with the (010) plane being the first plane to disappear (Fig. 2d). In the crystal structure of MIOC-7 (Fig. 2e), the (101) plane corresponds to molecular correlations within the 1D chains, the (010) plane is related to the chain extension along the b axis, whereas the (001) plane is indicative of correlations between interdigitated layers. At the initial stage of melting (505 K), slippage of the 1D chains induces rapid disorder within the interdigitated layers. Upon further heating (508 K), the strong hydrogen bonds are weakened, leading to the progressive disintegration of the chains and increasing disorder between the interdigitated layers. Ultimately (510 K), the strong hydrogen bonds break, intermolecular correlations collapse into complete disorder, and the melt is formed (Fig. 2f).

The results indicate that the melting of MIOC crystals begins with the slippage and distortion of strong hydrogen-bonded structures, followed by rapid disordering upon the breaking of the hydrogen bonds. As the starting point of the glass formation, the sacrifice of hydrogen bonds provides the free volume required for molecular mobility, thereby avoiding both the breaking of coordination bonds and ligand decomposition. The stacked hydrogen-bonded network structure (MIOC-4) and the interdigitated structure enriched with multiple π-π interactions (MIOC-7), constructed through ligand selection, exhibit distinct thermal responses. The hydrogen-bonded network readily undergoes slippage and distortion, whereas the interdigitated structure provides additional stability and dissociates progressively with increasing molecular thermal motion. This underscores the role of structural tailoring in controlling the thermal properties of the products.

**Structural analyses**

To understand the structure changes caused by vitrification, we analyze the FT-IR and Raman spectra of the MIOC crystals and glasses (Fig. 3). Here, we choose $ZnCl_2(HIm_{0.123}HbIm_{0.877})_2$ mixed crystals as a representative sample of the binary system. In FT-IR spectra (Fig. 3a), the peaks at 1065 $cm^{-1}$ are seen, assigned to the C-H in-plane bending mode of HIm for both MIOC-4 crystal and glass[26]. The peaks at 433 and 422 $cm^{-1}$ are ascribed to the in-plane ring bend mode of HbIm in MIOC-7 crystal and glass[27]. Both mixed crystals and their glassy counterpart exhibit the FT-IR signals of HIm and HbIm. The sharp peaks around 3306 and 700 $cm^{-1}$ for the MIOC crystals are associated with the stretching mode of N-H bonds from N-H···Cl and bending mode of C-H bonds from C-H···Cl, respectively[20, 28]. The peaks of N-H bonds in all the samples become broadened and exhibit red-shift upon vitrification, while the peaks of the C-H bonds diminish. This implies that glass formation weakens the C-H···Cl hydrogen bonds while strengthening the N-H···Cl hydrogen bonds. Based on the structural evolution observed during the melting of MIOC crystals, a schematic of hydrogen-bond environment changes during glass formation is shown in Fig. 3b. In the MIOC crystals, the break of N-H···Cl and C-H···Cl hydrogen bonds drives melting. During quenching, N-H···Cl hydrogen bonds reform and dominate the local structure and hydrogen-bonded network. The change of the hydrogen bonding environment is attributed to the fact that the bonding energy of N-H···Cl bonds is higher than that of C-H···Cl bonds. Therefore, the formation of the disordered N-H···Cl hydrogen bond network occurs during the liquid-to-glass transition[20],[29]. The Raman spectra (Fig. 3c) of all the samples show a peak at 299 cm$^{-1}$, which is attributed to the stretching modes of Zn-Cl, confirming the presence of the Cl ligand before and after vitrification[30]. Compared to the crystals, the glasses display weakened and broadened lattice peaks, likely arising from subtle short-range disordering of the tetrahedra[31]. Interestingly, the MIOC-7 glass shows a stronger fluorescence background compared to other crystalline and glassy samples, which is ascribed to the impact of compositional selection (HbIm in MIOC-7) on the luminescent properties of the complex.

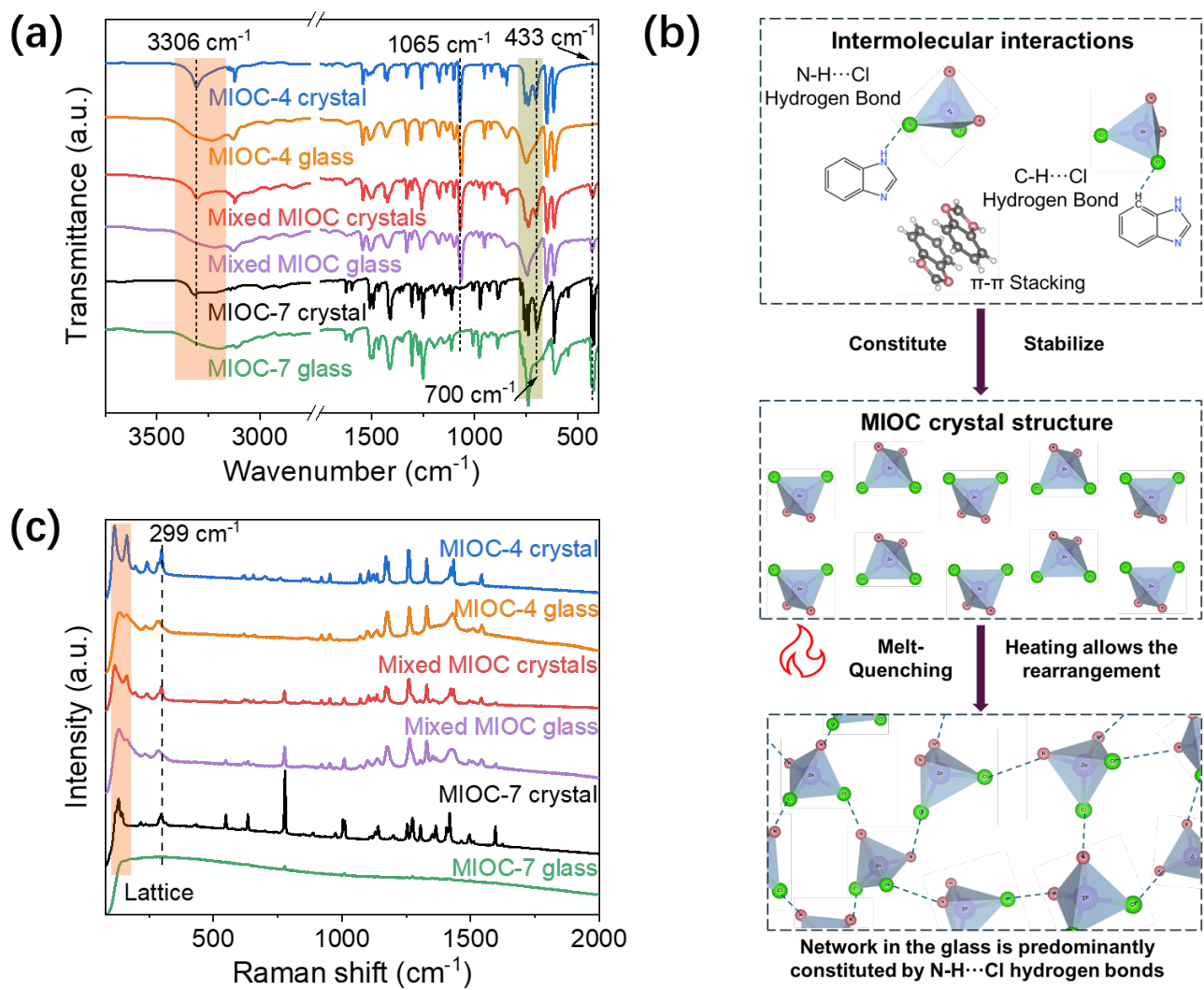

**Figure 3. (a)** Fourier transform infrared (FT-IR) spectroscopy of MIOC crystals and glasses. **(b)** Schematic diagram of the MIOC glass formation mechanism. **(c)** Raman spectra of MIOC crystals and glasses.

Hereafter, we focus on the MIOC-7 because only MIOC-7 glass ($T_g$ = 360 K) could be ground into a powder and remain stable at room temperature. To reveal the short-range structural evolution in MIOC crystals during vitrification, we conduct $^1$H, $^{13}$C, $^{15}$N, and $^{67}$Zn magic angle spinning (MAS) nuclear magnetic resonance (NMR) characterizations on MIOC-7 crystal and glass. Figs. 4a-c and S29 show the normalized $^1$H, $^{15}$N, $^{67}$Zn, and $^{13}$C NMR spectra of MIOC-7 crystal and glass, respectively. In the $^1$H NMR spectrum (Fig. 4a), MIOC-7 shows two main resonance peaks at about 7.2 and 8.1 ppm and three weak ones at 1.9, 4.0, and 12.7 ppm, whereas MIOC-7 glass shows one main resonance peak at 7.2 ppm and one weak peak at 12.4 ppm. The $^1$H NMR spectrum of MIOC-7 is deconvoluted into five resonance peaks (Fig. S27). However, peaks at 1.9 and 4.0 ppm vanish, and the peak at 8.1 ppm shifts to lower frequency after vitrification (Fig. S28). Specifically, the 1.9 ppm resonance is attributed to residual acetone trapped within the crystal lattice[32]. Its disappearance after vitrification agrees with the DSC results that exhibit the desolvation of trapped acetone (Fig. 1e). The 4.0 ppm signal corresponds to H species stemming from the moisture in atmosphere[33], which disappear upon desolvation. The two main resonances at 8.1 ppm and 7.2 ppm arise from H atoms bonded to C atoms on the aromatic ring. The 8.1 ppm resonance is attributed to H atoms of the imidazole ring part and those H atoms engaged in the C-H···Cl bond because this interaction lowers electron density of the affected H atoms. Upon vitrification, the 8.1 ppm resonance shifts to 7.3 ppm in the MIOC-7 crystal, indicating the disappearance of the C-H···Cl hydrogen bond. The 12.7 ppm resonance is ascribed to H atoms bonded to N atoms[34]. This phenomenon can be explained as follows. The C-H···Cl hydrogen bonds are replaced by disordered N-H···Cl hydrogen bonds upon vitrification, resulting in the formation of both more uniform C-H bonds and a wider variety of N-H···Cl hydrogen bonding networks.

Fig. S29 shows the $^{13}$C NMR spectra of the MIOC-7 crystal and glass. It is observed that upon vitrification, their $^{13}$C NMR peaks only slightly move to a lower chemical shift (see figure inset). This indicates that the deshielding effect around the C atoms in the MIOC-7 crystal becomes weaker, i.e., vitrification can increase the electron density around C. The $^{15}$N NMR spectra of MIOC-7 crystal and glass (Fig. 4b) exhibit peaks at 154 and 197 ppm, which are attributed to the N atoms bonded to H and the N atoms coordinated to Zn, respectively[34, 35]. In the crystal, the signals appear as pairs at 153/159 and 191/199 ppm, whereas they merge into two broad singlets in the glass. This can be attributed to the presence of two types of HbIm ligands in the crystal, one forming stacked planar layers and the other forming interdigitated layers, giving rise to the chemical shift differences. By contrast, the merging and broadening of peaks in the glass suggest a wide distribution of bond lengths and bond angles within the tetrahedra, reflecting a subtle short-range disorder. Interestingly, similar to ZIF glasses, the $^{67}$Zn NMR spectrum (Fig. 4c) of MIOC-7 glass exhibits a pronounced asymmetric peak shape and low-frequency tail relative to its crystalline counterpart. This further confirms short-range structural disorder in the MIOC glass[31] and suggests the freezing of multiple conformations of tetrahedral units. The hydrogen-bonded network is soft and incapable of forming rigid, ordered local structures. In the glass, molecular packing combined with the steric hindrance of HbIm limits the ability of molecules to return to their equilibrium positions (i.e., short-range order structure).

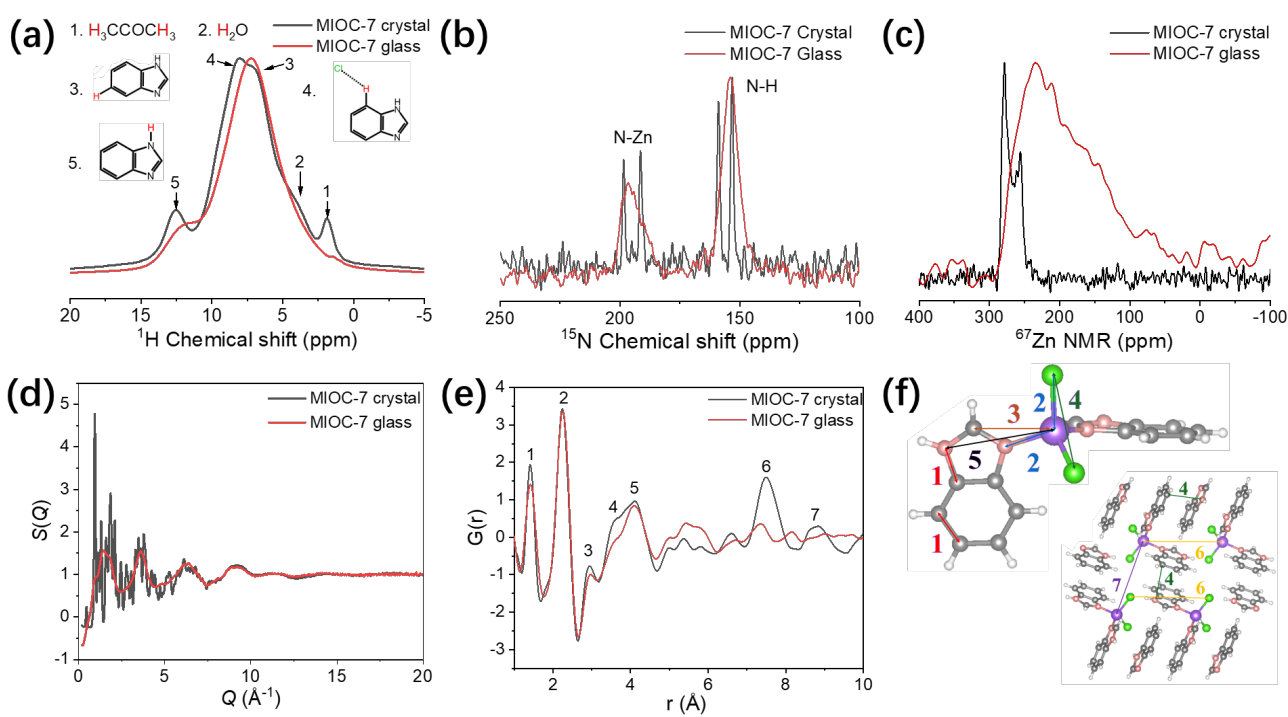

**Figure 4.** (a) Normalized $^1$H Solid-state magic-angle spinning (MAS) nuclear magnetic resonance (NMR) spectra of MIOC-7 crystal and glass. (b) Normalized $^{15}$N cross-polarization MAS (CPMAS) NMR spectra of MIOC-7 crystal and glass. (c) Normalized $^{67}$Zn CPMAS NMR spectra of MIOC-7 crystal and glass. (d) Faber-Ziman structure factor $S(Q)$ of MIOC-7 crystal and glass. (e) Reduced pair correlation functions $G(r)$ of MIOC-7 crystal and glass. (f) Schematic diagram of the atomic relationships in the MIOC-7 structure. Noting: purple spheres represent Zn, gray spheres represent C, green spheres represent Cl, white spheres represent H, and pink spheres represent N.

High-energy X-ray diffraction technique is employed to probe the short-, medium-, and long-range structural evolution of MIOC-7 crystal upon vitrification. The acquired diffraction images collected by synchrotron XRD are processed using the self-made igor program[36]. Fig. 4d shows the Faber-Ziman structure factor $S(Q)$ curve of both MIOC-7 crystal and glass, where sharp diffraction peaks are present in the crystal, whereas broader peaks appear in the glass. This further confirms that the long-range order vanishes upon vitrification[37]. The $G(r)$ curves in Fig. 4e show seven main atom-atom correlation peaks in MIOC-7 crystal, and five main peaks in MIOC-7 glass. The atom-atom correlations are schematically displayed in Fig. 4f[38-41]. Peak 1 at 1.4 Å corresponds to C-C(N) interaction in the HbIm rings. Peak 2 at 2.2 Å is ascribed to the overlap of the correlations of both the nearest-neighbor Zn-N (2.0 Å) and Zn-Cl (2.2 Å), implying that the metal-ligand bonding remains intact in MIOC-7 glass. Peak 3 at 2.95 Å and peak 5 at 4.1 Å are attributed to Zn-C and Zn-N (next-neighbored N) correlations, respectively. These intramolecular correlations remain unchanged upon glass formation. Peak 4 at 3.6 Å is assigned to both the nearest-neighbored Cl-Cl correlation and the aromatic ring-ring correlations. Considering no change of peak 2 (Zn-Cl correlation), the intensity decrease of peak 4 upon glass formation indicates that the π-π stacking is disrupted. Peak 6 at 7.5 Å corresponds to intermolecular nearest-neighbored Zn-Zn and Cl-Cl correlations, while peak 7 at 8.8 Å is ascribed to the next-neighbored Zn-Zn correlations. The reduced pair correlation functions $G(r)$ curves demonstrate that the Zn-ligand correlations remain intact in the MIOC-7 glass since the C-C(N), Zn-N, and Zn-Cl correlation peaks in MIOC-7 glass closely match those of MIOC-7 crystal. In contrast, the intermolecular correlations vanish in MIOC-7 glass, suggesting that the hydrogen-bonded network becomes disordered in the medium-range.

## Extension of the ligand-tuning approach

The ligand-tuning approach is further extended through multilevel structural tuning to synthesize four $MX_2L_2$ crystals, where M is a tetracoordinate-favored metal node (Co, Cu), X is a hydrogen bond acceptor (Cl⁻ or SCN⁻), and L is a hydrogen bond donor (HIm or HbIm). Figs. S30-33 show XRD patterns for four MIOC crystals with distinct hydrogen-bonding networks and crystal structures, confirming their successful synthesis[14, 42-44]. Figs. S34-37 present DSC upscan curves for four MIOC crystals, where the $T_m$ and $T_g$ values are determined (Table S3). The results indicate that metal nodes exert only minor effects on the thermal properties of MIOCs, whereas variations in hydrogen-bond strength and network topology are the key determinants of their thermal behavior. These results confirm the universality of our approach to designing hydrogen-bonded network glasses.

## Optical properties of MIOC-7 crystal and glass

Based on the observation of the strong fluorescence of MIOC-7 glass (Fig. 3c), it is interesting to examine whether the obtained glasses have unique photonic properties due to the ligand selection. Therefore, we characterize the optical properties of MIOC-7 crystal and glass. Fig. 5a shows the optical absorption spectrum of MIOC-7 glass, where a sharp increase of the absorbance with decreasing wavelength is observed at around 400 nm. The absorption signal corresponds to the π electron transition from the valence to the conduction band of the HbIm. The absorption edge is determined to be 350 nm from the increased absorbance, from which the band gap is estimated to be about 3.5 eV. This value is consistent with that calculated by periodic density functional theory (PDFT)[45].

Figs. 5b and c show the two-dimensional photoluminescence (excitation-emission) spectra of MIOC-7 crystal and glass, respectively, which are obtained from their excitation and PL emission spectra (Figs. S38 and S39). One narrow emission peak is seen at 308 nm in the crystal under excitation in the wavelength range of 250-290 nm. Three emission peaks are observed at 336, 353, and 371 nm in the crystal under excitation in the wavelength range of 300-330 nm. MIOC-7 glass shows three broad emission peaks at 373, 392, and 412 nm under excitation in 300-370 nm and a broad emission peak in the visible range under excitation in 370-550 nm. Interestingly, the broad emission peaks occurring in the glass under excitation in 370-550 nm shift with excitation wavelength towards longer wavelength, i.e., leading to a red shift. The structural origin of the observed photoluminescence (PL) behaviors can be explained as follows and illustrated in Fig. 5d. First, the narrow emission peak at 308 nm corresponds to the periodically arranged HbIm, where the rigidity of the crystal suppresses the nonradiative decay. Second, the three emission peaks both in the crystal and glass are attributed to the π-π* electron transition of HbIm[46, 47]. The broadening and red shift of the PL peaks of the crystal are ascribed to the distortion of the medium- and long-range structure. Third, the excitation-dependent PL of the glass might be associated with the giant red edge effect caused by slow relaxation[48].

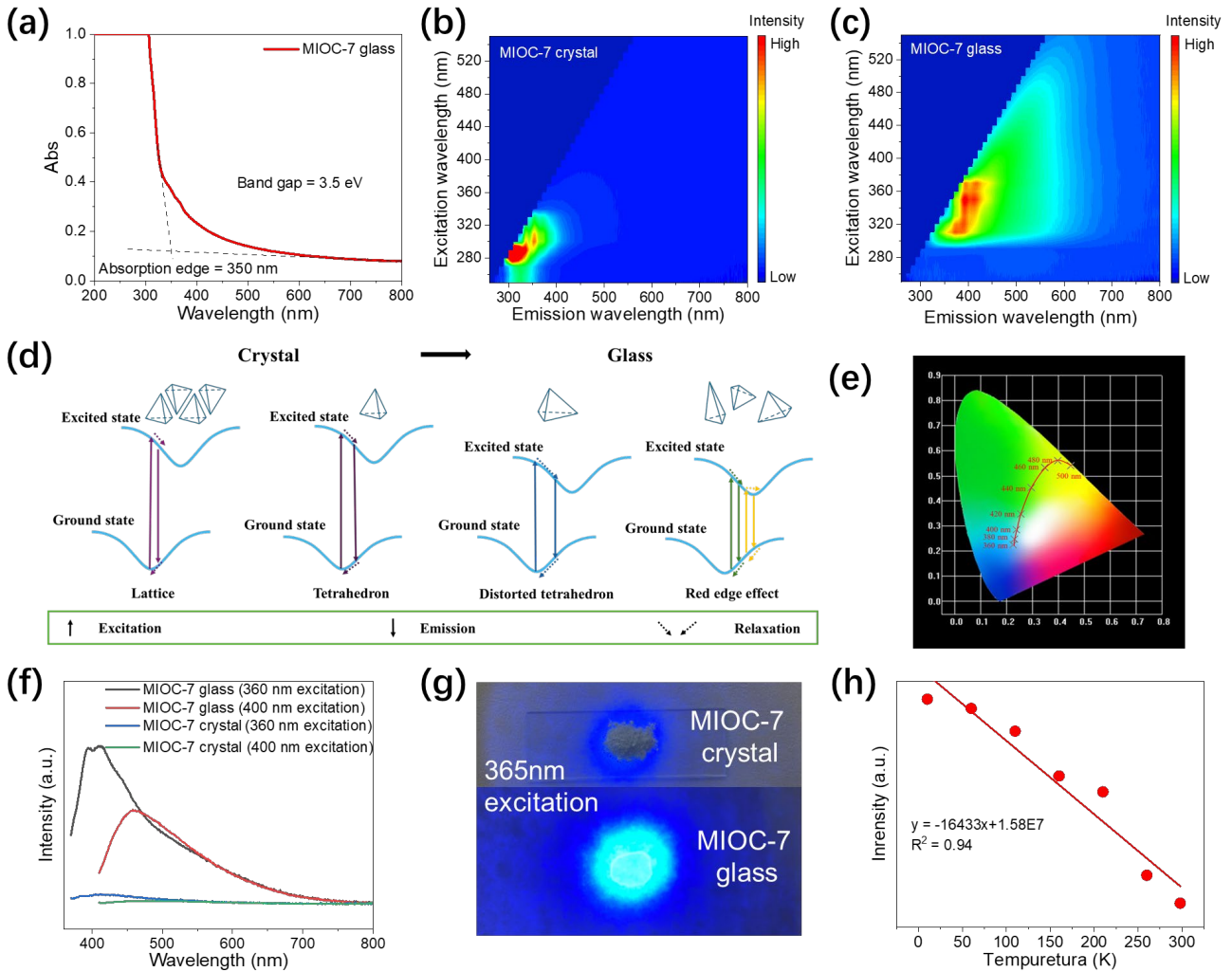

**Figure 5. (a)** Absorption spectrum of MIOC-7 glass. **(b)** Two-dimensional fluorescence (Excitation/Emission) spectra of MIOC-7 crystal. **(c)** Two-dimensional fluorescence (excitation-emission) spectra of MIOC-7 glass. **(d)** Schematic diagram of the luminescence mechanism. **(e)** Commission on Illumination (CIE) chromaticity coordinates of the emission spectra of MIOC-7 glass. **(f)** Fluorescence spectra of MIOC-7 crystal and glass under 360 nm and 400 nm excitation. **(g)** Photos of MIOC-7 crystal and glass under excitation of a 365 nm laser. **(h)** Relationship between temperature and photoluminescence intensity of MIOC-7 glass under 360 nm excitation.

The excitation-dependent PL color points of the MIOC-7 glass are marked in International Commission on Illumination (CIE) chromaticity coordinates (Fig. 5e), demonstrating that the PL color can be altered from blue to yellow by increasing the excitation wavelength. This anti-Kasha behavior is attributed to the red-edge effect arising from slow molecular relaxation in the rigid glass system[48]. Distinct conformers of $ZnCl_2HbIm_2$ molecules are frozen within the glass network, as evidenced by the broadening of the $^{67}Zn$ NMR signal of MIOC-7 crystals upon vitrification (Fig. 4c). Different excitation lights selectively excite molecular subpopulations with distinct energies, while the hydrogen bond network hinders conformational transitions, resulting in fluorescence emission from non-equilibrated states and thus a red-shifted emission[49]. Compared to MIOC-7 crystal, MIOC-7 glass exhibits significantly enhanced PL under the excitation at 360 and 400 nm (Fig. 5f). Photographs of MIOC-7 crystal (top) and glass (bottom) under excitation at 365 nm are shown in Fig. 5g, clearly exhibiting the strong blue light emission of MIOC-7 glass. In addition, the Internal PL quantum yield (PLQY) of MIOC-7 glass (3.4%) is around three times higher than that of MIOC-7 crystal (1.2%) under excitation at 370 nm (Table. S4). By integrating the PL intensity over the wavelength, the integrated value of MIOC-7 glass linearly decreases with an increase in temperature due to the increased probability of non-radiative transition (Figs. 5h and S40). This remarkable PL of the MIOC-7 glass demonstrates the potential for anti-counterfeiting applications.

## Conclusions

We designed two hydrogen-bonded MIOC crystals and glasses with the composition of $ZnCl_2HIm_2$ and $ZnCl_2HbIm_2$ by a ligand-tuning approach. The hydrogen-bonded network is established by substituting the coordination bonds of Im/bIm ligands in ZIFs with hydrogen bonds introduced by $Cl^-$ ligand. Upon heating, structural differences governed distinct phase evolution processes. Stacked structure in MIOC-4 promoted slippage and distortion of hydrogen-bonded chains, whereas interdigitated structure in MIOC-7 preserved the chains until complete disorder, enhancing thermal stability. The breaking of hydrogen bonds finally induces melting. Upon quenching, the medium- and long-range intermolecular correlations of both Zn-Zn and Cl-Cl vanish. Meanwhile, short-range correlations such as Zn-N, Zn-Cl, and Zn-C are preserved in the glass but are frozen into the network in a short-range disordered state.

Mixed-ligand synthesis enables the formation of binary MIOC mixed-crystals. In the mixed-crystal system, defects such as dislocations and mixed ligand molecules resulted in a decrease in the melting point. The lowest melting point in the system is 415 K for the $ZnCl_2HbIm_{0.498}HIm_{1.502}$ mixed crystals. After melt-quenching, $T_g$ increases from 282 to 360 K upon gradually substituting HIm by HbIm. This indicates that the steric hindrance of tetrahedral units can be influenced by composition selection, enabling linear control of the $T_g$ of MIOC glasses. In addition, we expanded the ligand-tuning strategy to include metal node-tuning (Zn, Co, and Cu) and acid anion ligand-tuning (SCN$^-$). Finally, we successfully obtained a series of MIOC crystals and glasses, demonstrating the potential of the ligand-tuning approach for expanding the glass-forming MIOC family.

After vitrification of MIOC-7 crystals, we observed enhanced photoluminescence (PL) in the visible light region under excitation at the wavelength of 365 nm. In addition, the internal PL quantum yield (PLQY) of MIOC-7 glass (3.4%) is nearly three times higher than that of MIOC-7 crystal (1.2%). The excitation-dependent PL in the glass suggests that [Zn-N/Cl] tetrahedra with distinct conformations are frozen in the hydrogen-bonded network, reflecting the connection between the short-range disorder within the glass and its fluorescence properties. The remarkable PL of the MIOC-7 glass shows the potential for the application of anti-counterfeiting.

# Acknowledgments

The authors wish to thank the China Scholarship Council (grant 202208310046) for Tianzhao Xu's scholarship. Thanks to the proposal number 2024A1372 and Dr. Hiroki Yamada and Dr. Seiya Shimono (Japan Synchrotron Radiation Research Institute) for their help in HEXRD measurements and analysis. We also thank the State Key Laboratory of Silicate Materials for Architectures (SYSJJ2025-11).

# Competing interests

The authors declare that they have no known competing financial interests or personal relationships that could have appeared to influence the work reported in this paper.